\documentclass{ieeeaccess}
\usepackage{cite}
\usepackage{amsmath,amssymb,amsfonts}
\usepackage{algorithmic}
\usepackage{graphicx}
\usepackage{textcomp}
\usepackage{listings}
\usepackage{tikz}
\usetikzlibrary{matrix}
\NewSpotColorSpace{PANTONE}
\AddSpotColor{PANTONE} {PANTONE3015C} {PANTONE\SpotSpace 3015\SpotSpace C} {1 0.3 0 0.2}
\SetPageColorSpace{PANTONE}

\usepackage{bm}
\makeatletter
\AtBeginDocument{\DeclareMathVersion{bold}
\SetSymbolFont{operators}{bold}{T1}{times}{b}{n}
\SetSymbolFont{NewLetters}{bold}{T1}{times}{b}{it}
\SetMathAlphabet{\mathrm}{bold}{T1}{times}{b}{n}
\SetMathAlphabet{\mathit}{bold}{T1}{times}{b}{it}
\SetMathAlphabet{\mathbf}{bold}{T1}{times}{b}{n}
\SetMathAlphabet{\mathtt}{bold}{OT1}{pcr}{b}{n}
\SetSymbolFont{symbols}{bold}{OMS}{cmsy}{b}{n}
\renewcommand\boldmath{\@nomath\boldmath\mathversion{bold}}}
\makeatother

\def\BibTeX{{\rm B\kern-.05em{\sc i\kern-.025em b}\kern-.08em
    T\kern-.1667em\lower.7ex\hbox{E}\kern-.125emX}}

\begin{document}
\history{Date of publication xxxx 00, 0000, date of current version xxxx 00, 0000.}
\doi{10.1109/ACCESS.2024.0429000}

\title{Comments on: RIO: Return Instruction Obfuscation for Bare-Metal IoT Devices with Binary Analysis}
\author{\uppercase{Kai Lehniger}\authorrefmark{1},
\uppercase{Peter Langendörfer}\authorrefmark{1}\authorrefmark{2}}

\address[1]{IHP - Leibniz-Institut f\"ur innovative Mikroelektronik, 15236 Frankfurt (Oder), Germany (\{lehniger, langendoerfer\}@ihp-microelectronics.com)}
\address[2]{Brandenburgische Technische Universität Cottbus-Senftenberg, 03046 Cottbus, Germany (e-mail: peter.langendoerfer@b-tu.de)}

\markboth
{Lehniger \headeretal: Comments on: RIO: Return Instruction Obfuscation for Bare-Metal IoT Devices with Binary Analysis}
{Lehniger \headeretal: Comments on: RIO: Return Instruction Obfuscation for Bare-Metal IoT Devices with Binary Analysis}

\corresp{Corresponding author: Kai Lehniger (e-mail: lehniger@ihp-microelectronics.com).}

\begin{abstract}
This is a comment on "RIO: Return Instruction Obfuscation for Bare-Metal IoT Devices with Binary Analysis". RIO prevents finding gadgets for Return-Oriented Programming attacks by encrypting return instructions. This paper shows flaws in the design of RIO that allow for the easy retrieval of the plaintext return instructions without decrypting them. Additionally, changes are proposed to improve upon the original idea.
\end{abstract}

\titlepgskip=-21pt

\maketitle

\section{Introduction}

Return-oriented programming (ROP) \cite{shacham2007geometry} is a popular attack method for applications written in unsafe languages. By utilizing some kind of memory vulnerability, an attacker can overwrite a return address with his payload, jumping to a gadget, a small code snippet that ends with a return instruction. Each return of a gadget will pop the next gadget address from the stack, creating a gadget chain that allows arbitrary computations.

Several countermeasures against ROP attacks have been proposed, one of it being Return Instruction Obfuscation (RIO) \cite{kim2023rio}. The basic presumption RIO works with is that an attacker needs to know the return instructions in order to find gadgets with binary analysis. Without gadgets, no ROP attack can be performed. This paper will describe a method which allows to recover the unencrypted return instruction with high precision without the need of breaking the encryption. Afterwards, possible improvements will be discussed.

The rest of this paper is structured as follows. Section II briefly summarizes the function return mechanic in ARM, Section III gives an overview of RIO, and Section IV discusses possibilities to break its protection. Section V shows possible improvements of the original design and Section VI concludes the paper.

\section{Function Returns in ARM}

ARM processors have a dedicated register that holds return addresses, \texttt{lr}, the link register. When \texttt{bl} or \texttt{blx} instructions are used to perform a function call, the return address is automatically being put in the \texttt{lr} register. A function return can be performed with \texttt{bx lr}. In such a case, the return address is never exposed to an attacker that has access to the stack. 
When more function calls happen, \texttt{lr} needs to be stored in the stack. Instead of restoring its value in the end, the return address can be popped directly from the stack into the program counter. This is typically done together with other register values in a single \texttt{pop \{reglist\}} instruction. Attackers can use this by controlling not only the return address but also register values that can serve as an input for the gadget.

\section{RIO: Return Instruction Obfuscation}

RIO \cite{kim2023rio} removes the possibility to find gadgets by encrypting all return instructions. Typically, gadget finding algorithms start to search for gadgets with the return instruction. By removing this starting point from the binary, these algorithms need to be adapted in order to work again. Also, by removing the information which registers are popped from the stack, it is no longer possible to construct a payload, as it is unknown how much the stack pointer advances and where in the payload to place the next gadget address.

RIO is implemented by using compiler custom passes of the LLVM compiler. An initialization module is inserted in the beginning of the existing firmware and a Return Control Flow (RCF) module is placed in front of every return instruction \cite{kim2023rio}. The return instructions themselves are encrypted. When running the firmware, the initialization module scans the binary for all RCF modules and decrypts subsequent return instructions. The decrypted instructions are placed in a table in SRAM.

The RCF module consists of three instructions shown in \lstlistingname~\ref{lst:rcf}. The first instruction loads the base address of the table into register \texttt{r0}, which is placed after the encrypted return instruction. The second instruction adds an offset to the base address to point to the corresponding return instruction in the table. The last instruction jumps into the table, where the return instruction is executed.
\begin{lstlisting}[float,label=lst:rcf, caption=Return Control Flow module code \cite{kim2023rio}]
ldr     r0, [pc, #12]
adds    r0, #offset
mov     pc, r0
\end{lstlisting}

\section{Attack Points}

RIO has three major flaws that attackers could exploit, two of them being used to adapt binary analysis in order to find gadgets.

The first flaw is the general assumption that in order to perform ROP attacks, an attacker needs to perform binary analysis. There have been examples in the past of attacking a device without having a copy of the target binary \cite{bittau2014hacking}. However, without binary analysis, ROP attacks become more difficult and therefore the goal of RIO to hinder this step is still worth pursuing.

The second flaw is the assumption that RIO hides the position of return instructions. The RIO initialization module itself uses the RCF modules in order to find the positions of return instructions. Consequently, any attacker could do the same. Of course only knowing the position of return instructions is not enough for ARM, as the exact registers need to be known in order to prepare the payload.

This can be done by using the last flaw of the approach: RIO keeps the \texttt{push} instructions in the prologue unencrypted. Typically, function prologues and epilogues are symmetrical. Registers that are being pushed onto the stack in the prologue are being popped in the epilogue. This can be illustrated by taking the code examples from \cite{kim2023rio} that were used to demonstrate the encryption. \lstlistingname~\ref{lst:examples} shows the \texttt{push} and \texttt{pop} instructions in the prologue and epilogue of two functions. In the first function \textit{RIOtest} four registers \texttt{r4}, \texttt{r6}, \texttt{r7}, and \texttt{lr} are pushed onto the stack in the prologue. In the epilogue, the same registers are popped, with the only difference of \texttt{lr} being replaced with \texttt{pc}. The same pattern repeats for \textit{main2} with the registers \texttt{r7} and \texttt{lr}. Even when completely removing the \texttt{pop} instructions from the binary, they could easily be derived by looking at what registers where pushed onto the stack and need to be restored. Of course, instead of restoring \texttt{lr}, the return address is directly popped into \texttt{pc}.

\begin{lstlisting}[float,label=lst:examples, caption=Symmetrical push and pop instructions of unencrypted functions taken from \cite{kim2023rio}]
RIOtest:
    push    {r4, r6, r7, lr}
    ...
    pop     {r4, r6, r7, pc}

main2:
    push    {r7, lr}
    ...
    pop     {r7, pc}
\end{lstlisting}

\section{Possible Improvements}

While the first two flaws are inherent and cannot be targeted, as the encrypted return addresses must be locatable in order to decrypt them, the third flaw can be addressed. By extending the encryption to the \texttt{push} instructions as well, using a similar method, the information of what registers are being stored on the stack can be hidden. Doing so would take a bit more effort due to the fact that, after the execution of the decrypted \texttt{push} instruction, the control flow needs to jump back to the function.

However, even encrypting both, \texttt{push} and \texttt{pop}, instructions would probably not suffice. The reason to push and pop registers to the stack in the prologue and epilogue is given by the calling conventions. Callee saved registers are registers that must remain unchanged for the caller when calling a function. In order to meet this guarantee, the callee is required to store and restore all callee saved registers it uses during its execution. Consequently, by analysing which callee saved registers are being used inside a function, the \texttt{push} and \texttt{pop} instructions can be derived.

\begin{figure*}
    \centering
    \begin{tikzpicture}[row 1/.append style={nodes={draw=none,minimum width=0mm}},
                        row 2/.append style={nodes={draw=none,minimum width=0mm}},
                        row 3/.append style={nodes={draw=none,minimum width=0mm}}]
        \matrix[matrix of nodes, column sep=5mm, nodes={draw, minimum width=30mm, font=\ttfamily,right},nodes in empty cells]{
            pop \{ r4, r6, r7, pc\} & pop \{r6, r7, lr\}    & pop \{r7, lr\}    & pop \{lr\}\\
                                    & pop \{r4\}            & pop \{r4, r6\}    & pop \{r4, r6, r7\}\\
                                    & bx lr                 & bx lr             & bx lr\\
            |[draw=none]|\\
            r4                      & r6                    & r7                & |(b)|ret addr\\
            r6                      & r7                    & ret addr          & r4\\
            r7                      & ret addr              & r4                & r6\\
            ret addr                & r4                    & r6                & |(a)|r7\\
        };
        \draw[->] ([xshift=5mm]a.south east) -- node[rotate=90,below]{stack growth} ([xshift=5mm]b.north east);
    \end{tikzpicture}
    \caption{Example return instruction and the corresponding layout in the stack (left) alongside possible replacements with different positions of return addresses}
    \label{fig:ret_replacements}
\end{figure*}
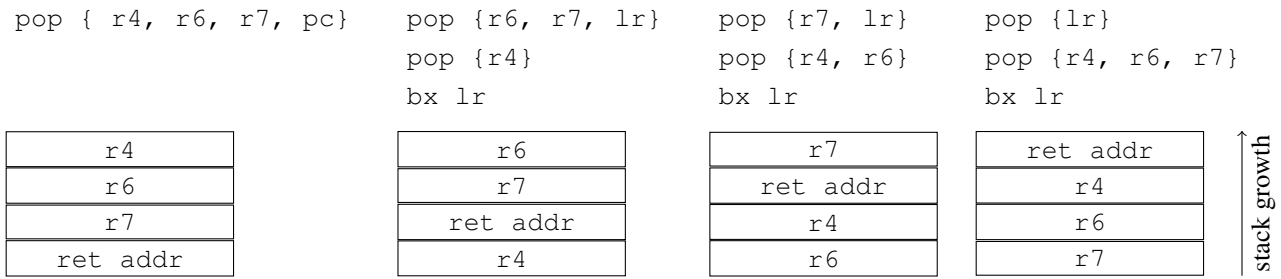

The authors found two ways to remove this method of deriving the instructions that can be used in conjunction:
\begin{itemize}
    \item For each \texttt{push} and \texttt{pop} pair a random number of additional registers can be added. These additional registers would not be possible to be derived from binary analysis.
    \item Each decrypted \texttt{push} and \texttt{pop} pair in the table could be replaced by multiple instructions to randomize the position of the return addresses in each stack frame. Since the table is created at runtime, the positions on the stack would be different with each reset of the IoT device. \figurename~\ref{fig:ret_replacements} shows a return instruction on the left, as well as possible replacement sequences next to it and how it effects the position of the return address in the stack. The corresponding \texttt{push} instructions have to be changed in a similar way to match the layout.
\end{itemize}

\section{Conclusion}

This paper showed several severe flaws of the RIO implementation and presented ideas how to use these flaws to gain knowledge of the encrypted return instructions without knowledge of the secret key. Additionally, improvements have been proposed to remove possibilities of binary analysis for attackers.

\bibliographystyle{IEEEtran}
\bibliography{literature}

\begin{IEEEbiography}[{\includegraphics[width=1in,height=1.25in,clip,keepaspectratio]{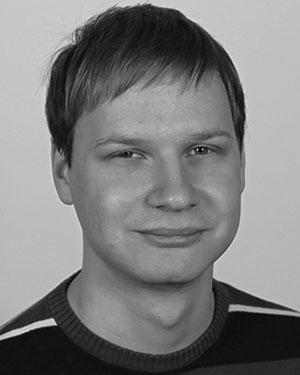}}]{Kai Lehniger}
received the master’s degree in computer science, in 2017. Since 2017 he is with the IHP in Frankfurt (Oder). There, he is working in the wireless systems department as a scientist. He has published more than 10 papers. He currently is interested in efficient security for resource constrained devices.
\end{IEEEbiography}

\begin{IEEEbiography}[{\includegraphics[width=1in,height=1.25in,clip,keepaspectratio]{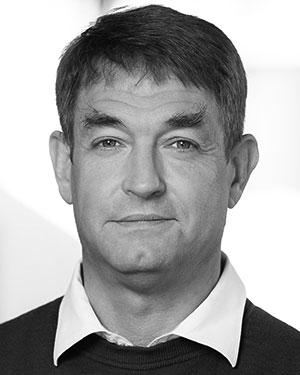}}]{Peter Langendörfer}
received the diploma, in 1995, and the doctorate degree in computer science, in 2001. Since 2000 he is with the IHP in Frankfurt (Oder). There, he is leading the Wireless Systems Department. From 2012 till 2020 he was leading the chair for security in pervasive systems with the Technical University of Cottbus-Senftenberg. Since 2020 he owns the chair wireless systems with the Technical University of Cottbus-Senftenberg. He has published more than 150 refereed technical articles, filed 17 patents of which 11 have been granted already. He is associate editor of IEEE Access, IEEE Internet of Things, Peer-to-Peer Networking and worked as guest editor for many renowned journals e.g., Wireless Communications and Mobile Computing (Wiley) and ACM Transactions on Internet Technology. He is highly interested in security for resource constraint devices, low power protocols, efficient implementations of AI means and resilience. He is member of the “Gesellschaft für Informatik.”
\end{IEEEbiography}

\EOD

\end{document}